\documentclass[twocolumn,aps,showpacs,preprintnumbers,nofootinbib,
superscriiptaddress]{revtex4}
\usepackage{graphicx}
\usepackage{dcolumn}
\usepackage{bm}
\begin{document}

\title{On violation of the Robinson's damping criterion and enhanced
cooling of ion, electron and muon beams in storage rings }

\author{E.G.Bessonov, Lebedev Physical Institute RAS, Moscow, Russia}

\date{\today}

                       \begin{abstract}
Limits of applicability of the Robinson's damping criterion \cite{1958}
and the problem of enhanced cooling of particle beams in storage rings
beyond the criterion are discussed.  \end{abstract}


\maketitle

\section{Introduction}

Losses of energy by particles in storage rings caused by friction
forces can lead to damping of their betatron and synchrotron
oscillations and to a decrease of the six-dimensional phase space
volume occupied by particles in a beam (cooling of a beam). Friction
can be caused by:  1) radiation reaction in static external fields, 2)
radiation reaction in time dependent external fields (backward Compton
and Rayleigh scattering of laser light), 3) ionization and radiation
processes in material targets located in the vacuum chamber of the
ring. Special insertion devices can be used to introduce friction
forces. The physics of damping in the longitudinal plane is different
from that in the transverse one.

Usually a particle with a higher than ideal energy loses more energy
than the ideal particle and a particle with lower energy loses less
energy.  The combined effect is that the energy difference between
non-ideal and ideal particles is reduced. A decrease of the particle
energy deviation from the ideal leads to a decrease of its longitudinal
phase deviation from the synchronous phase.

In the transverse plane, the loss of particle energy leads to a loss of
its longitudinal as well as transverse momentum, since the particle
performs betatron oscillations. The total momentum loss is, however,
replaced in the cavity only in the longitudinal direction. The combined
effect of the energy loss and the replacement of the energy loss in
accelerating cavities leads to a net loss of transverse momentum.

The total amount of damping in all degrees of freedom is limited by the
amount of the energy loss.  There is a correlation of damping
decrements of different planes determined by Robinson's damping
criterion \cite{1958}-\cite {bruck}.  This criterion limits the rate of
particle cooling in storage rings.  Below, we would like to pay
attention to the limits of applicability of this criterion, its
violation in schemes of selective interaction of particle beams with
targets.

We start by reviewing Robinson's damping criterion.

\section{ LIMITS OF APPLICABILITY OF THE ROBINSON'S DAMPING CRITERION}

The motion of particles in storage rings is described by their
deviations from the ideal orbit in transverse radial $x$, vertical $y$
directions and by deviation $\varphi = \psi - \psi _s$ of the particle
phase from the synchronous one in a curvilinear coordinate system
($x,y,\varphi $). In a linear approximation deviations (x,y,$\varphi $)
are described by linear second order differential equations.  We can
write the equations in the form of a system of six linear first order
differential equations for a six-dimensional coordinate vector $\vec u
$ with components ($x,x{'},y,y{'},\varphi ,\Delta \varepsilon $), where
$x{'}=\partial x/\partial s$; $y{'}=\partial y/\partial s$; $s$, the
longitudinal coordinate of a particle along the ideal (reference)
orbit; $\Delta \varepsilon = \varepsilon - \varepsilon _s$, the
deviation of the particle energy from synchronous one. In the matrix
presentation:

\begin{equation}                     
\label{eq1}
\frac{d\vec u (s)}{ds} = Q(s)\vec u (s),
\end{equation}
where $Q(s) = || q_{ji} (s) ||$ is a six-order matrix with components
$q_{ji} (s)$ ($j$,$i$ = 1...6). The Eq.(\ref{eq1}) has six linear
independent solutions $\vec {u_j } (s)$ with components $u_{ji} (s)$
($u_{j1} = x_j$, $u_{j2}=x{'}_j $, $u_{j3}=y_j $,... $u_{j6}=\Delta
\varepsilon _j $). The solution of the Eq. (\ref{eq1}) has the form
$\vec u (s) = U(s) \cdot \vec u (0)$, where $U(s) = ||u_{ij} (s)||$ is
a transfer matrix; $\vec u (0)$, the initial vector.  The determinant
of this matrix is a Wronskian $W(s) = |U(s)|$. It represents the
six-dimensional volume of the polyhedron in the phase space occupied by
the beam. The values $dW(s) / ds = SpQ \cdot W(s)$,\hskip 2mm $W(s) =
W(0) \cdot \exp (\int SpQ \cdot ds) \simeq W(0) \cdot \exp ( < SpQ >
\cdot s)$, where $SpQ = \sum\nolimits_{j = 1}^6 {q _{jj} } $, $W(0)$ is
the initial Wronskian, sign $ < > $ denotes averaging. This is the
Jacobean formula, \cite{ 1951}. On the other hand,
$u_{ji} (s) \sim \exp ( \alpha _i \cdot s)$ and the rate of change of
the 6-dimensional volume of the polyhedron $ \sim \exp [ 2\sum {\alpha
_i } s]$, where $\alpha _i=\alpha _x ,\alpha _y,\alpha _{\varepsilon} $
are fractional averaged damping decrements.  Therefore

      \begin{equation}    
      \label{eq2}
      \alpha _{6D} = 2 \sum \limits_{i = 1}^3 {\alpha _i } = < SpQ > .
      \end{equation}

$SpQ$ is determined by the diagonal elements of the matrix $\vert \vert
q_{ji} (s)\vert \vert $.  In the transverse plane, the particle
momentum loss does not lead to a change of the direction of the
momentum and position of the particle ($q _{11} = q _{33} = q _{55}
=0$).  Acceleration of the particle changes the direction of the
momentum on the value $\vert \Delta \vec {p\vert } / \vert \vec p
\vert = \overline {P _s} \cdot s / c \cdot \varepsilon _s $. It leads to
matrix elements $q_{22} = q_{44} = - \overline {P _s} / c \cdot
\varepsilon _s $, where $ \overline {P _s}$ is the average rate of
particle energy loss; $\varepsilon $, the particle energy; subscript
$s$ refers to the reference orbit; $c$, the velocity of light. The rate
of change of the particle energy is $\partial \varepsilon/ \partial t =
- (\partial \overline {P} / \partial \varepsilon )|_s \cdot \Delta
\varepsilon  + (\partial P_{rf} / \partial \psi )\vert _{s } \cdot
\varphi $ and matrix element $q_{66} = - (\partial \overline {P} / c
\cdot \partial \varepsilon )|_s $.  Substitution of diagonal matrix
elements to (\ref{eq2}) leads to generalized Robinson damping criterion
\cite {kolom2}, \cite{wiedemann}:

\begin{equation} 
\label{eq3}
\sum \limits_{i = 1}^3 {\alpha _i} = \frac{1}{2} \alpha _{6D} =
- \frac{1}{c}{\frac{ \overline {P _s} }{\varepsilon _s}
- \frac{1}{2 c} \frac{\partial \overline {P } }{ \partial
\varepsilon }|_s} .  \end{equation}

The proof of the Robinson's damping criterion was reduced to
application mathematical Jacobean formula. Non-diagonal matrix
components responsible for the beam dynamics of particles in a lattice
was not used. Two diagonal components responsible for damping in the
transverse plane are determined by the average power of the particle
energy loss and one diagonal component responsible for damping in the
longitudinal plane is determined by the partial derivative of the power
energy loss. The value $\alpha _{6D}$ in (\ref{eq2})
determine the rate of damping of the 6-dimensional phase space volume
(emittance) occupied by the beam (cooling). Coefficients $\alpha _{\varepsilon}$
and $\alpha _{6D}$ can be both positive and negative \cite {kolom},
\cite {kolom2}.

If there is no coupling between radial $x$ and vertical $y$ planes in a
storage ring, the direct calculations can be performed separately for
vertical and longitudinal damping coefficients:

\begin{equation}   
\label{eq4}
\alpha _y = - \frac{1}{2c}\frac{ \overline {P _s} }{\varepsilon _s },
\hskip 10mm
\alpha _{\varepsilon} = - \frac{1}{2c}\frac{d \overline {P} }
{d\varepsilon}|_s .
\end{equation}

The radial decrement follows from Eq. (\ref{eq3}):

\begin{equation}       
\label{eq5}
\alpha _x = - \frac{1}{2c}\left[ {\frac{ \overline {P _s} }{\varepsilon
_s} + \frac{\partial \overline {P} }{\partial \varepsilon }|_s - \frac{d
\overline {P}}{d\varepsilon }|_s} \right].
\end{equation}

Damping times $\tau _{i} = - 1/c\alpha _{i}$ are limited by Robinson's
damping criterion (\ref{eq3}) by the value

     \begin{equation}                
     \label{eq6}
     \tau _{i} = \frac{\varepsilon _s}{J_{i}\overline {P _s}}.
     \end{equation}
where $J_{i}$ is determined by the dependence \hbox { $\overline
P(\varepsilon)$}. All decrements must be $J_{i} \geq 0$. Every
decrement $J _{i} \leq J _{i, max} = 2$ if radiation in external fields
is emitted (synchrotron, undulator, backward Compton scattering one,
$\overline P \sim \varepsilon ^2$). In case of radiative ion cooling
based on backward Rayleigh scattering in the homogeneous laser beam
having uniform spectral distribution the value $J _{i} \leq J _{i, max}
= (3+2D)/2(1+D)$, where $D$ is the saturation parameter ($\overline P
\sim D/(1 + D)$, $D \sim \varepsilon$) \cite{prl}. For ionization muon
cooling $({\partial \overline {P} }/ { \partial \varepsilon })|_s$ is
rapidly decreasing with the energy for $\varepsilon _{\mu} < 0.3$ GeV,
but is slightly increasing for $\varepsilon _{\mu} > 0.3$ GeV ($J _{i,
max} \sim 1$) \cite{1983}, \cite {neufer}. The partial derivative
$({\partial \overline {P} }/ { \partial \varepsilon })|_s$ can be very
high only in case of laser cooling of ion beams by homogeneous
broadband laser beam with rapidly increasing linear dependant spectral
intensity in the frequency range corresponding to the ion energy spread
$\sigma _{\varepsilon, \,0}$. In this case $\overline P/ \overline P
_{s}  \simeq (\varepsilon - \varepsilon _s + \sigma _{\varepsilon,
\,0})/ \sigma _{\varepsilon, \,0}$ ($-\sigma _{\varepsilon, \,0} <
\varepsilon - \varepsilon _s < \sigma _{\varepsilon, \,0}$),
$({\partial \overline {P} }/ { \partial \varepsilon })|_s = \overline P
_{s} / \sigma _{\varepsilon, \,0} \gg \overline P _s /\varepsilon _s$,
$J _{i, max} \simeq \varepsilon _s/ 2 \sigma _{\varepsilon, \, 0}$,

              \begin{equation}                
             \label{eq7}
             \tau _{\varepsilon} = \frac{2 \sigma
             _{\varepsilon, \,0}} { \overline P _{s}}.  \end{equation}

The value $\tau _{6D} = - 1/c\alpha _{6D}$  determine the damping time
of the six-dimensional phase space volume of the beam. In case of
radiation of particles in external fields the value $\tau _{6D} > 0$
(cooling). At the same time in this case the damping in one plane and
antidamping in another one is possible \cite {kolom}. In case of
ionization losses of energy by muons at small energies the damping time
of the six-dimensional phase space volume of the beam $\tau _{6D} <
0$ (heating) \cite {kolom2}.

Next conditions were used to prove the Robinson's damping criterion:
1) cooling in the radio frequency (RF) bucket, 2) linearity of the
system 3) stationary conditions. Violation of these conditions can lead
to the violation of the criterion, the concept of decrement
(non-exponential damping) and fast cooling.

Below we consider schemes of enhanced emittance exchange and
six-dimensional enhanced cooling of particle beams beyond conditions
used to prove the Robinson's damping criterion.

The term enhanced we refer to the emittance exchange and cooling times
determined approximately by the equation (\ref{eq7}) and less. During
this time the energy losses of particles of a beam are about the energy
spread of the beam.

\section{Enhanced damping schemes based on selective interaction of
particle beams with moving targets}

In this section two schemes of enhanced damping of particle beams in
the longitudinal and transverse planes are considered when external
selective interaction of particles with moving targets is used \cite
{ICFA00} - \cite {heacc01}.  The targets (material, laser beam,
undulator) are installed in the vacuum chamber of the storage ring and
can be displaced in the radial direction. The RF acceleration system is
switched off and there is no energy dependence of the average rate of
the particle energy loss. Material targets can be used for muon beams
and laser targets can be used for ion and electron beams (laser beams
propagate in the opposite direction to ion or electron beams)\footnote
{Stabilization schemes allow parallel displacement of the laser beam
relative to the ion beam to better than 10 $\mu$m \cite{lauer}. In
reality, closed orbits can be moved in the direction of the target
instead of moving the target. A kick, decreasing of the magnetic field
in bending magnets of the storage ring, a phase displacement or eddy
electric fields can be used for this purpose.}.  The interaction region
(IR) in this case changes its radial position in the particle beam
during the damping process.  Broadband laser beams have to be used for
ion beam damping in order for all ions, independent of their energy, to
interact with laser beams.

The enhanced damping schemes are presented in Figures 1 and 2. In these
Figures the axis $"z"$ is the unwrapped reference orbit of the storage
ring, $L_l$ and $a$ are the length and the width of the targets in the
interaction region, respectively.  The transverse positions of targets
are displaced along the axis "x" with the velocities $v _{T _{1}} > 0$
and $v _{T _{2}} < 0$ relative to the reference orbit. 1,2,3,.. are the
particle trajectories before and after the energy loss in the target.
\hbox {$O_1$, $O_2$, $O_3$, .  .  .} are the locations of the particle
closed orbit after \hbox {0,1,2,3, .  . .} energy loss events. Parts of
one period of the particle trajectory at the targets $T _1$ and $T _2$
are shown before and after every event of the particle energy loss in
the targets.

\begin{figure}[hbt]
\includegraphics{
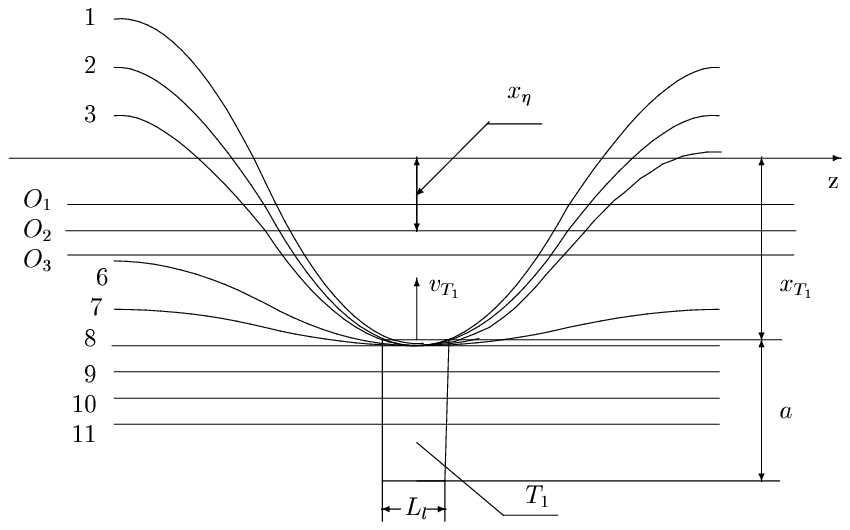}
\caption{
Scheme of enhanced damping of a particle beam in transverse plane. The
evolution of the amplitude of betatron oscillations is shown for the
target at rest.} \end{figure}

A target $T _1$ can be used for damping of betatron oscillation
amplitudes of particle beams (see Fig. 1). At the initial moment it
overlaps a small internal part of the particle beam in the radial
direction in the straight section of a storage ring with non-zero
dispersion function.  The degree of overlapping of particle beam and
target is changed by moving the target position uniformly with some
velocity $v _{T _1}$ from inside in the direction of the particle beam.
First, particles with largest initial amplitudes of betatron
oscillations interact with the counter-propagating target. Immediately
after the interaction and loss of energy, the position and direction of
momentum of a particle remain the same, but the closed orbit is
displaced inward in the direction of the target \cite {wiedemann}. The
radial coordinate of the closed orbit and the amplitude of betatron
oscillations are decreased by the same value owing to the dispersion
coupling. After every interaction, the position of the closed orbit
approaches the target more and more, and the amplitude of betatron
oscillations is reduced.

The closed orbit moves with a velocity  $\dot x_{\eta}$, which depends
on the particle's amplitude of betatron oscillation and the distance
between the orbit and the target, since these values determine the
probability of collision of a particle and a target (see below). It
reaches a maximum velocity $|\dot x_{\eta \, in}|$ ($\dot x_{\eta \,
in} < 0$) if the depth of penetration of the particle closed orbit in
the target is greater than the amplitude of betatron oscillations of
this particle\footnote {The value $\dot x_{\eta \, in} = \delta x
_{\eta}/T <0$, $T = 1/f$ is the revolution period.}. After this, the
amplitude of betatron oscillations will stay constant since the
particle will interact with the homogeneous target every turn for both
positive and negative deviation from the orbit with equal probability.
When the target reaches the closed orbit corresponding to particles of
maximum energy and the depth of penetration of the particle closed
orbits in the target is greater than the amplitude of betatron
oscillations of these particles, it must be returned to its previous
position. All particles of the beam will have small amplitudes of
betatron oscillations and increased energy spread.

Particles with high amplitudes of betatron oscillations start
interacting with the target first. Their interaction time is longer
and hence the decrease of amplitudes of betatron oscillations is
greater.

\hspace{15mm}
\begin{figure}[hbt]
\includegraphics{
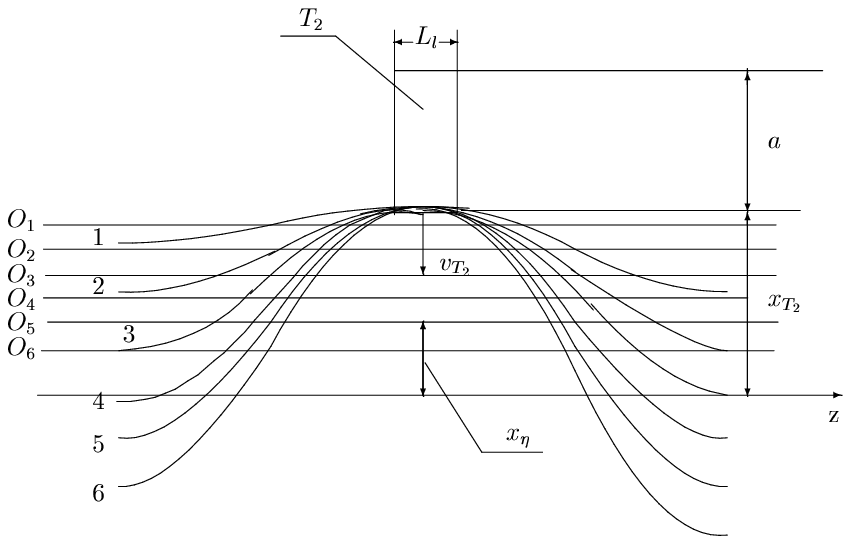}
\caption{
Scheme of enhanced damping of a particle beam in longitudinal plane.
The evolution of the amplitude of betatron oscillations is shown for
the target at rest.} \end{figure}


A target $T _2$ can be used for damping of the energy spread of
particle beams (see Fig. 2). The radial target position must be moved
uniformly with a velocity  $|v _{T _2}| > |\dot x_{\eta \, in}|$ from
outside in the direction of the particle beam being cooled. At the
initial moment, the target overlaps only a small part of the particle
beam. The degree of overlapping is changed in such a way that particles
of maximum energy first come into interaction and then particles of
lower energy. When the target reaches the orbit of particles of minimum
energy, it must be switched off and returned to the previous
position. In this case the difference in duration of interaction and
hence in the energy losses of particles having maximum and minimum
energies will be large.  As a result, all particles will be gathered at
the minimum energy in a short time.

The evolution of the amplitude of betatron oscillations and position of
the closed orbit of a particle is determined by its energy loss in the
target. Contrary to the previous case the energy loss of the
particle leads to increase of the amplitude. But in this case the
radial velocity of the target displacement $|v _{T _2}| > |\dot
x_{\eta \, in}|$ and hence after a short time the particle closed orbit
will be deepened into the target to a depth greater than the
amplitude of its betatron oscillation and the increase will end.

             \subsection {Theory of interaction of particle beams with
             transversely moving targets}

Below we will neglect the energy dependence of the average rate of
the particle energy loss in the target, the emission of SR by particles
in the bending magnets of a storage ring, assuming the RF system of the
ring is switched off, targets are homogeneous and have sharp edges in
the radial directions.  We assume the jump of closed orbits of
particles caused by their energy loss in the target is much less than
amplitudes of betatron oscillations.

In a smooth approximation, the motion of a particle relative to the
position $x_{\eta}$ of its closed orbit is described by the equation $x
_{\beta} = $ $A _0 \cos \varphi _{\beta}$, where $x _{\beta} = x -
x_{\eta}$ is the particle deviation from the orbit; $x$, its radial
coordinate; $\varphi _{\beta}= \omega _{\beta}(t - t _0) + \varphi
_{\beta 0}$, phase of the particle betatron oscillations; $\varphi
_{\beta 0}$ and $t _0$, the initial phase and time; $A _0$ and $\omega
_{\beta}= 2\pi \nu _x f$, the initial amplitude and the frequency of
betatron oscillations, respectively; $\nu _x$, the tune of betatron
oscillations; $f$, the particle revolution frequency \cite {wiedemann}.
If the coordinate $x _{\beta\,0}$ and transverse radial velocity of the
particle $\dot x _{\beta\,0} = - A _0 \omega _{\beta} \sin \varphi
_{\beta 0}$ correspond to the moment $t _0$ of change of the particle
energy in a target, then the amplitude of betatron oscillations of the
particle before an interaction is $A _0$ $ =$ $ \sqrt {x _{\beta\,0} ^2
+ \dot x _{\beta \,0} ^2 /\omega _{\beta} ^2}$. After the interaction,
the position of the particle closed orbit will be changed by a value
$\delta x_{\eta} = \eta _x \beta ^ {-2}(\delta {\varepsilon} /
{\varepsilon})$, where  $\eta _x$ is the dispersion function of the
storage ring; $\beta$, the relative velocity of the particle; $\delta
{\varepsilon}$, the change of the particle energy. The deviation of the
particle relative to the new orbit will be $x _{\beta \,0} - \delta x
_{\eta}$, and the direction of the particle velocity will not be
changed. The new amplitude will be $A_1$ $ =\sqrt {(x _{\beta \,0} -
\delta x _{\eta})^2 + \dot x ^2 _{\beta \,0} /\omega _{\beta} ^2}$ and
the change of the square of the amplitude

       \begin{equation}
       \label{eq8}
       \delta A ^2 = A_1^2 - A_0^2 = - 2x _{\beta \,0}\delta x _{\eta}
       + (\delta x _{\eta}) ^2.
       \end{equation}          

In the approximation $|\delta x _{\eta}| \ll |x _{\beta\,0}| < A _0$,
the value

       \begin{equation}         
             \label{eq9}
       \delta A = - {x _{\beta \,0} \over A} \delta x _{\eta}.
       \end{equation}

The rate of change of the particle betatron oscillation amplitude is
maximal when $|x _{\beta 0}| = A$.

The velocity of a particle orbit $\dot x_{\eta}$ depends on the
probability $W$ of collision of particle and target in one turn. This
probability is determined by the distance $x _{T _{1,2}} - x _{\eta}$
between the edge of the target and the particle closed orbit and on the
amplitude of betatron oscillations. Particles interact with the target
at every turn ($W=1$) and their radial velocity reaches the maximum
value $\dot x _{\eta \, in}$ when the orbit enters the target at a
depth larger than the amplitude of betatron oscillations. In the
general case $\dot x_{\eta} = W \cdot \dot x_{\eta \, in}$.

The deviation of a particle from its closed orbit at the location of
the target takes on values $x _{\beta n} = A \cos \varphi _{\beta n}$,
where $\varphi _{\beta n} = 2\pi \nu _x n + \varphi _{\beta 0}$,
$n=1,2,3,$... . Different values $x _{\beta n}$ in the region ($-A, A$)
will occur with equal probability if $\nu _x \simeq p/q$, $q \gg 1$
(tune is far from forbidden resonances), $p$ and $q$ are integers.  The
collision of targets and particle beams occur when $x _{\beta n} \leq x
_{T _1} - x _{\eta} $ and $x _{\beta n} \geq x _{T _2} - x _{\eta} $.
 These conditions are valid when $\varphi _{\beta n}$ are in the range
of phases $2\varphi _{T _{1,\,2}}$, where $\varphi _{T _{1}} = \pi -
\arccos \xi _{1}$ and $\varphi _{T _{2}} = \arccos \xi _{2}$, $\xi
_{1,\,2} = (x _{T _{1,\,2}} - x _{\eta})/A$. Particles cross through the
target at the range of phases $2\varphi _{T_{1,2}}$ and cross over it
at the range of phases $2\pi - 2\varphi _{T_{1,2}}$.  The probability
can be presented in the form $W = \varphi _{T_{1,2}}/\pi$ and the value
$\dot x_{\eta} = \varphi _{T_{1,2}} \cdot \dot x_{\eta \, in} /\pi $,
where $\dot x_{\eta \, in} = - \eta _x \beta ^ {-2} (\overline P /
{\varepsilon})$.

The behavior of amplitudes of betatron oscillations of particles is
determined by (\ref{eq9}).  Particles cross the target at different
coordinates $x _{\beta\,0}$ in the range of phases $2\varphi
_{T_{1,2}}$. That is why the average rate of change of amplitudes
$\partial A/\partial x _{\eta} = $ $-\overline x _{\beta\,0}/A$, where
$-\overline x _{\beta\,0}$ $ = -(1/ \varphi_{T_{1,2}})\int _{a
_{1,2}} ^{b _{1,2}} A \cos \varphi _{\beta n} d \varphi _{\beta n} =
\pm A sinc \, \varphi _{T_{1 ,2}}$ and $sinc \,\varphi _{T_{1,2}}$ $ =
$ $ sin \varphi _{T_{1,2}}/\varphi _{T_{1,2}}$; the signs $+$ and $-$
are related to the first and second targets. We took into account that
the limits of integration $a _{1,2}$ and $b _{1,2}$ depend on the
target: $a _{1} = \arccos \xi _1 = \pi - \varphi _{T_{1}}$, $a _{2} = 0
$, $b _{1} = \pi$, $b _{2} = \varphi _{T_{2}}$.

Thus, the evolution of amplitudes and closed orbits is determined by
the system of equations

       \begin{equation}               
             \label{eq10}
       {\partial A\over \partial x_{\eta}} = \pm sinc \,\varphi
       _{T_{1,2}}, \hskip 5mm {\partial x _{\eta}\over \partial t} =
       {\dot x_{\eta \, in} \over \pi}\varphi _{T_{1,2}}.
       \end{equation}

From equations (\ref{eq10}) and the expression $\partial A/\partial x
_{\eta} = [\partial A/\partial t]/ [\partial x _{\eta}/\partial t]$, it
follows:

       \begin{equation}                                
             \label{eq11}
       {\partial A\over \partial t} = {\dot x _{\eta \,in}\over
       \pi}\sin\varphi _{T_{1,2}} = {\dot x _{\eta \,in}\over
       \pi}\sqrt {1 - \xi _{1,\,2}^2}.  \end{equation}

Let the initial closed particle orbits be distributed in a region $\pm
\sigma _{\eta, 0}$ relative to the location of the middle closed orbit
and the initial amplitudes of particle radial betatron oscillations $A
_0$ be distributed in a region $\sigma _{x,0}$ relative to their closed
orbits. The spread of closed orbits $\sigma _{\eta, 0} = |\eta _x|
\beta ^ {-2}$ $(\sigma _ {\varepsilon,0} / {\varepsilon})$.

Suppose that the initial spread of amplitudes of betatron oscillations
of particles $\sigma _{x,0}$ is identical for all closed orbits of the
beam.  The velocities of the closed orbits in a target $\dot x
_{\eta \, in} < 0$, the transverse velocities of the targets $v _{T
_{1}} > 0$ and $v _{T _{2}} < 0$. Below, we will use the relative
radial velocities of the target displacement $k _{1,2} = v _{T
_{1,2}}/\dot x _{\eta \, in}$, where $v _{T _{1,2}} = dx _{T
_{1,2}}/dt$.  In our case, $\dot x_{\eta \, in} < 0$, $k _1 < 0$, $k _2
> 1$.

From the definition of $\xi _{1, 2}$ we have the relation $x _{\eta} =
x _{T _{1,2}} - \xi _{1, 2} A(\xi _{1, 2})$. The time derivative is
$\partial x _{\eta}/\partial t = v _{T _{1,2}} - [A + \xi _{1,2}
(\partial A/ \partial \xi _{1,2})] \partial \xi _{1,2} /\partial t$.
Equating this value to the second term in (\ref{eq10}), we obtain the time
derivative

        \begin{equation}                               
             \label{eq12}
        {\partial \xi _{1,2}\over \partial t} = {\dot x _{\eta \,in}
        \over \pi} {\pi k _{1,2} - \varphi _{T_{1,2}} \over A(\xi
        _{1,2}) + \xi _{1,2} (\partial A / \partial \xi
        _{1,2})}.  \end{equation}

Using this equation we can transform the first value in (\ref{eq10}) to
the form $\pm sinc \varphi _{T_{1,2}} (\xi _{1,2}) = ({\partial A/
\partial \xi _{1, 2}})$ $({\partial \xi _{1,2}/ \partial t})/$ $
({\partial x _{\eta}/ \partial t}) = $ $(\pi k _{1,2} - \varphi
_{T_{1,2}}) {(\partial A /\partial \xi _{1,2})}/[A + \xi
_{1,2}(\partial A/\partial \xi _{1,2})]\cdot \varphi _{T_{1,2}}$
or $\partial \ln A / \partial \xi _{1,2}$ $ = \pm
\sin \varphi _{T_{1,2}} /$ $[\pi k _{1,2} - (\varphi _{T_{1,2}} \pm \xi
_{1,2} \sin \varphi _{T_{1,2}})]$. Substitution $Z = \varphi
_{T_{1,2}}\pm \xi _{1,2} \sin \varphi _{T_{1,2}}$ leads to the
solution of the last equation

       $$A = A _0 \exp \int _{\xi _{1,2,0}} ^   {\xi _{1,2}}
       {\pm \sin \varphi _{T_{1,2}} d\xi _{1,2}
       \over \pi k _{1,2} - (\varphi _{T_{1,2}}\pm \xi _{1,2}
       \sin \varphi _{T_{1,2}})} $$
       \begin{equation}                         
             \label{eq13}
       = A _0 \sqrt {\pi k _{1,2} - Z(\xi _{1,2,0})
       \over \pi k _{1,2} - Z(\xi _{1,2}) } = A _0 \sqrt {\pi k _{1,2}
       \over \pi k _{1,2} - Z(\xi _{1,2}) }, \end{equation}
where the index $0$ corresponds to the initial time $t _0$, when $\xi
_{1,0} = -1, \xi _{2,0} = 1$. Substitution $A$ and $\partial A/
\partial \xi _{1,2}$ in (\ref{eq12}) leads to the time dependence $\xi
_{1,2}(t)$ in the form

       \begin{equation}                                
             \label{eq14}
       t - t _0 = {\pi A _0 \over |\dot x _{\eta \,in}|} |\psi
       (k _{1,2}, \xi _{1,2})|,   \end{equation}
where $\psi (k _{1,2}, \xi _{1,2}) = - \int ^{\xi _{1,2 }} _{\xi _{1,2,
0}} d\xi _{1,2 } A(\xi _{1,2 })/ A_0 [ \pi k _{1, 2} - (\varphi
_{T_{1,2}}$ $ \pm \xi _{1,2 }\sin \varphi _{T_{1,2}})]$ or, according
to (\ref{eq13}),

       $$\psi (k _{1,2}, \xi _{1,2}) = \int _{\xi _{1,2,0}} ^{\xi
       _{1,2}}{- \sqrt{\pi k _{1,2}} d\,\xi _{1,2} \over |\pi k _{1,
       2} - (\varphi _{T_{1,2}} \pm \xi _{1,2 }\sin \varphi _{T
       _{1,2}})| ^{3/2}} $$
       \begin{equation}             
             \label{eq15}
       = \pm \int _{\xi _{1,2,0}} ^{\xi
       _{1,2}}{\sqrt{\pi |k _{1,2}|} d\,\xi _{1,2} \over |\pi k _{1, 2}
        - (\varphi _{T_{1,2}} \pm \xi _{1,2 }\sin \varphi _{T _{1,2}})|
       ^{3/2}}.  \end{equation}

Closed orbits of particles having initial amplitudes of betatron
oscillations $A _0$, according to (14), penetrate into the target
to a depth greater than their final amplitudes of betatron oscillations
$A _f$ at a moment $t _f = t _0 + $ $\pi A _0 \psi (k _{1,2}, \xi
_{1,2,f})/$ $|\dot x _{\eta \, in}|$, where $\xi _{1,f} = \xi _1(t _f)
= 1$, $\xi _{2,f} = \xi _2(t _f) = -1$. During the interval $t _f - t
_0$, the targets $T _{1,2}$ will cross a distance $|\Delta x _{T
_{1,2,f}}| = |v _{T _{1,2}}|(t _f - t _0) = \pi |k_{1,2}| |\psi (k
_{1,2}, \xi _{1,2,f})| A _0$.

The ratio of the final to the initial amplitude of betatron
oscillations, according to (\ref{eq13}), is

           \begin{equation}          
             \label{eq18}
           {A _f\over A _0} =
           \sqrt {k _{1,2} \over k _{1,2} - 1}.  \end{equation}

If $|\xi _{1,2}| < 1$, the position of the closed orbit, according to the
definition of $\xi _{1,2}$, can be presented in the form

          \begin{equation}      
          \label{eq16} x_{\eta}(t) = x_{T _{1, 2, 0}} + v _{T_{1,2}}(t
          - t_0) - A[(\xi _{1, 2} (t)] \cdot \xi _{1,2}(t).
          \end{equation}

At the moment $t _f$ the position of the closed orbit of a particle,
according to (\ref {eq16}), will be determined by the equation

             \begin {equation}  
             \label{eq19}
             x _{\eta, f} ^{'} = x _{\eta, 0} - \Psi (k _{1,2}) A
             _{0}, \end {equation}
where $\Psi (k _{1,2}) = - \xi _{1,2,0} + \xi _{1,2,f} (A _f /A _0) +
\pi k _{1,2} \psi (k _{1,2},$ $ \xi _{1,f})$. We have used a condition
$x _{T _{1,2,0}} = x _{\eta, 0} + A _0 \xi _{1,2,0}$.

The moment $t _0$ depends on the initial conditions for particles $A
_0$, $x _{\eta, 0}$ and targets. If the target has a position $x _{T
_{1,2,0}} = 0$ at a moment $t = 0$, then the target will contact particles
at the moment $t _0 = $ $x _0 / v _{T _{1,2}}$, where the position of the
particle $x _0 = x _{\eta,0} + A _0 \xi _{1,2,0}$ corresponds to the
minimal distance of the particle to the target at the moment $t = 0$. If
the last closed orbit of the beam will be deepened into the target to a
depth greater than all amplitudes of betatron oscillation of its particles
at a moment $t > t _{f,max} = max \{t  _f\}$, the orbits will be
distributed according to the law $x _{\eta}(t) = x _{\eta,f} ^{'} + \dot x
_{\eta, in} (t - t _f) = x _{\eta, 0} - \Psi (k _{1,2}) A _{0} + \dot x
_{\eta\,in} [t - \pi A _0 \psi (|k _{1,2}|, \xi _{1,f})/ |x_{\eta \,in}| -
(x _{\eta _0} + A _0 \xi _{1,2,0})\xi _{1,2,f} /v _{T_{1,2}}]$ or

\vskip -4mm

             $$x _{\eta,f} = x _{\eta,0}{k _{1,2} - 1\over k _{1,2}}
             + A _0 [\Psi (k _{1,2}) - \pi \psi (k _{1,2}, \xi
             _{1,2,f}) $$
             \begin {equation}  
             \label{eq20}
             +{\xi _{1,2,0}\over k _{1,2}}] + \dot x _{\eta\,in} \cdot t
             \hskip 6mm (t > t _{f,max}).
             \end {equation}

The duration of the cycle of the enhanced damping of particle betatron
oscillation amplitudes can be presented in the form $\tau _{x,
\varepsilon}|_{A_0 = \sigma _{x,0}} = 2\sigma _{\eta, 0}/ |v _{T _{1,2}}|
+ (t _f - t _0)$ or

         $$\tau _{x, \varepsilon} = {2\, \sigma _{\eta, 0}\over
         |k _{1,2}| |\dot x _{\eta \, in}|} + {\pi\, \sigma _{x,
         0} |\psi (k _{1,2}, \xi _{1,2,f})| \over |\dot x _{\eta \,
         in}|} $$
         \begin{equation}                
             \label{eq17}
         = {2\sigma _{\varepsilon,\,0} \over \overline P} \left [{1
         \over |k _{1,2}|} + {\pi \over 2 } R |\psi (k _{1,2}, \xi
          _{1,2,f})|\right],
         \end{equation}
where $R = {\sigma _{x,0} / \sigma _{\eta, 0}}$.

Dependencies $|\psi (k _{1,2}, \xi _{1,2, f})|$ and $\Psi (k _{1,2},
\xi _{1,2, f})$ determined by (\ref{eq15}), (\ref{eq19}) and the
dependence $|\psi (k _2, \xi _{2})|$ on $\xi _{2}$ for $k _2 = 1.0$, $k
_2 = 1.1$ and $k _2 = 1.5$ are presented in Tables 1 - 5, respectively.
The numerical values in these tables permit to analyze the evolution of
amplitudes of betatron oscillations and closed orbits in time.

\vskip 4mm
\begin{figure}[hbt]
Table 1. The dependencies $|\psi (k _1, \xi _{1, f})|$ and
$\Psi (k _1)$.
\hskip 20 mm \vskip 2mm
\begin{tabular}{|l|l|l|l|l|l|l|l|l|l|l|l|l}
\hline
$|k _1|$
&0.00& 0.01 & 0.02 & 0.04 & 0.08 & 0.16&0.32 &0.64&1.28&2.56&$\infty$\\
\hline
$|\psi |$
&$\infty$&2.42&1.93&1.54&1.21&0.94&0.70&0.50&0.32&0.19&0\\
\hline
$\Psi $
&1.00&$1.02$&1.02&1.00&0.97&0.90&0.79&0.63&0.45&0.28&0\\
\hline
\end{tabular}
\end{figure}

\vskip 4mm
\begin{figure}[hbt]
Table 2. The dependencies $|\psi (k _2, \xi _{2, f})|$ and
$\Psi (k _2)$.
\hskip 20 mm \vskip 2mm
\begin{tabular}{|l|l|l|l|l|l|l|l|l|l|l}
\hline
$k _2$
& 1.0 & 1.01 & 1.02 & 1.05 & 1.1&1.2 &1.4&1.8&2.6 \\
\hline
$|\psi |$
&$\infty$&24.38&13.8&6.52&3.71&2.10&1.18&0.65&0.35\\
\hline
$\Psi $
&$\infty$&66.3&36.1&16.0&8.50&4.49&2.34&1.20&0.61\\
\hline
\end{tabular}
\end{figure}

\vskip 4mm
\begin{figure}[hbt]
Table 3. The dependence $|\psi (k _2, \xi _{2})| _{k_2 = 1.0}$.
\vskip 2mm
\begin{tabular}{|l|l|l|l|l|l|l|l|l|l|l|}
\hline
$\xi _{2}$ & 1.0 & 0.5 & 0.2& 0&-0.2&-0.5&-0.8&-0.9& -1.0 \\
\hline
$|\psi |$ & $0$ &.182&.341&.492&.716& 1.393&4.388 &10.187&$ \infty $\\
\hline \end{tabular}

\vskip 4mm
Table 4. The dependence $|\psi (k _2, \xi _{2})| _{k_2 = 1.1}$.
\vskip 2mm
\begin{tabular}{|l|l|l|l|l|l|l|l|l|l|l|}
\hline
$\xi _{2}$ & 1.0 & 0.5 & 0.2&0 &-0.2&-0.5& -0.8 & -0.9& -1.0  \\
\hline
$|\psi |$&$0$&.163&.300&.423&.595&1.033& 2.076&2.759
& 3.710 \\
\hline \end{tabular}

\vskip 4mm
Table 5. The dependence $|\psi (k _2, \xi _{2})| _{k _2 = 1.5}$.
\vskip 2mm
\begin{tabular}{|l|l|l|l|l|l|l|l|l|l|l|l|l|l|}
\hline
$\xi _{2}$ & 1.0 & 0.5 & 0.2&0 & -0.2& -0.4&-0.6&-0.8&-1.0  \\
\hline
$|\psi |$&$0$&0.116&0.202&0.273&0.359& 0.466 &0.602 &0.772 &0.980 \\
\hline
\end{tabular}
\end{figure}

\subsection {Enhanced transverse damping}

In the method of enhanced damping of betatron oscillation amplitudes of
particles in a storage ring by a target $T _1$, the transverse betatron
beam size is decreased in accordance with (\ref{eq18}) to the value
$\sigma _{x,f}/ \sigma _{x,0} = A _f/ A _0 = \sqrt {|k _{1}|/ |(k _{1}|
+ 1)}$. The evolution of closed orbits of particles interacting with
the target depends on their initial amplitudes of betatron
oscillations.  First of all the target interacts with particles having
the largest initial amplitudes of betatron oscillations and the lowest
energies.  The position of a closed orbit of these particles is changed
by the law (\ref{eq16}) up to the moment $t _f$. For the time $t _f - t
_0$ the closed orbit will be displaced relative to its initial position
(or the position of particles having the same initial orbit but zero
amplitude of betatron oscillations) by the value (\ref{eq19}), where
$\Psi (|k _{1}|) = 1 + \sqrt {|k _1| / (1 + |k _1|)} - \pi |k _1| \psi
(|k _1|, \xi _{1,f})$.

According to (\ref{eq20}), the final relative dispersive beam size and
energy spread are

         $${\sigma _{\eta,f} \over \sigma _{\eta, 0}} = {\sigma
         _{\varepsilon, f} \over \sigma _{\varepsilon, 0}} =$$
         \begin{equation}                
             \label{eq21}
         {1 + |k _1| \over |k _1|} + 0.5 R [\Psi (k _1) -
         \pi \psi (k _{1}, \xi _{1,f}) + {1\over |k _1|}].
         \end{equation}

According to (\ref{eq18}), $\sigma _{x,f} / \sigma _{x,0} = 1/e \simeq
0.368$, if $|k _{1}| = |k _{1,e}| = 1/(e^2 - 1) \simeq 0.1565$.  In
this case

         \begin{equation}                
             \label{eq22}
         {\sigma _{\eta,f} \over \sigma _{\eta,0}}|_{R \ll 1} =
         {\sigma _{\varepsilon,f} \over \sigma
         _{\varepsilon,0}}|_{R \ll 1} \simeq 7.39.
         \end{equation}

The non-exponential damping time of the particle beam in the transverse
plane, according to (\ref{eq17}) and condition $k _1 = k _{1,e}$ can be
expressed in the form

          \begin{equation}                
          \label{eq24}
          \tau _{x}|_{R \ll 1} = {2\sigma _{\varepsilon,\,0} \over
          \overline P} [ e^2 - 1 ] \simeq
          {12.8 \sigma _{\varepsilon,\,0} \over \overline P}.
          \end{equation}

In this method of damping the width of the target

         \begin{equation}                
             \label{eq25}
         a > 2 (\sigma _{\eta,f} + \sigma _{x,f}).
         \end{equation}

         {\subsection {Enhanced longitudinal damping}

In the method of enhanced damping of the energy spread of particle
beams in storage rings by a target $T_2$, the transverse betatron beam
size is increased in accordance with (\ref{eq18}) to the value $\sigma
_{x,f}/ \sigma _{x,0} = A _f/ A _0 = \sqrt {k _{2}/ (k _{2} - 1)}$. The
evolution of closed orbits of particles interacting with the target
depends on their initial amplitudes of betatron oscillations. First of
all the target interacts with particles having the largest initial
amplitudes of betatron oscillations and the highest energies.  The
position of the closed orbit of these particles $x _{\eta _1}$ is
changed by law (\ref{eq16}) up to the time $t = t _f$. For the time $t
_f - t _0$ the closed orbit will be displaced relative to its initial
position (or the position of particles having the same initial orbit
but zero amplitude of betatron oscillations) by the value (\ref{eq19}),
where $\Psi (k _{2}) = \pi k _2 \psi (k _2, \xi _{2,f}) - \sqrt
{k _2 / (k _2 - 1)} - 1$.

According to (\ref{eq20}), the final relative dispersive beam size and
energy spread are

         $${\sigma _{\eta,f} \over \sigma _{\eta, 0}} = {\sigma
         _{\varepsilon, f} \over \sigma _{\varepsilon, 0}} =$$
         \begin{equation}                
             \label{eq26}
         {k _2 - 1 \over k _2} + 0.5 R [\Psi (k _2) -
         \pi \psi (k _{2}, \xi _{2,f}) + {1\over k _2}].
         \end{equation}

According to (\ref{eq18}), $\sigma _{x,f} / \sigma _{x,0} = \sqrt {e}
\simeq 1.65$, if $k _{2} = k _{2,e} = e/(e - 1) \simeq 1.58$. In this
case

         \begin{equation}                
             \label{eq27}
         {\sigma _{\varepsilon,f} \over \sigma _{\varepsilon,0}}
         |_{R \ll 1}
         \simeq {1 \over e} \simeq 0.37.
         \end{equation}

The non-exponential damping time of the particle beam in the
longitudinal plane, according to (\ref{eq17}) and condition $k _2 = k
_{2,e}$ can be expressed in the form

          \begin{equation}                
          \label{eq29}
          \tau _{\varepsilon}|_{R \ll 1} = {2\sigma _{\varepsilon,\,0}
          \over \overline P} {e - 1\over e} \simeq {1.27 \sigma
          _{\varepsilon,\,0} \over \overline P}.  \end{equation}

In this method of damping the width of the target is determined by
(\ref{eq25}) as well.

According to (\ref{eq18}), (\ref{eq20}), a rectangle on the plane ($x
_{\eta}, A$) is transformed to a parallelogram of the same area. The
four-dimensional phase space volume occupied by the beam in this case is
not changed in the geometrical variables ($x, x^{'}, y, y^{'}, s,
x_{\eta}$). In the canonical variables ($x, p_x, y, p_y, s, p_s$) the
Liouville's theorem does not work and we have a non-enhanced cooling
only\footnote{If we return particle beam to the initial energy state by
eddy accelerating fields, the transverse phase space volume in the
geometrical variables is decreased.}.

Friction and the external selectivity of interaction of homogeneous
moving targets with particle beams in storage rings do not lead to
enhanced cooling if the energy independent power loss in the target and
the approximation $|\delta x _{\eta}| \ll \sigma _{x,0}$ are
used\footnote {Terms "enhanced cooling in one plane" and "enhanced
heating in another one" used in our previous papers are misleading as
they concern to enhanced emittance exchange. The term "cooling" is
applied to damping of six-dimensional phase space volume determined in
canonical variables.}. In particular, with the framework of the above
approximation particles are deepened in the target $T _2$ to the depth
larger then their amplitudes of betatron oscillations for many turns,
interact with the target for this time at deviations from the closed
orbit $x _{\beta}$ of one sign and, according to (\ref{eq9}), receive
the unwanted increase of betatron amplitudes\footnote {This interaction
is similar to interaction of ions and monochromatic laser beams with
scanning frequency (see above). However, in the last case laser beams
overlap being cooled ion beams, interaction occur at the deviations $x
_{\beta }$ from their closed orbits of different signs, do not lead to
increase of amplitudes of betatron oscillations or leads to a weak
stochastic excitation of betatron oscillations if dispersion function
is not equal zero.}. If particles are deepened in the target to the
depth larger then their amplitudes of betatron oscillations for one
turn ($|\delta x _{\eta}| \geq \sigma _{x,0}$), they interact with the
target at deviations from their closed orbits of alternate signs.
However this and similar interactions do not lead to cooling. They lead
to change the location of the being interacted particles from one place
to another in the geometrical six-dimensional phase space without
change of the occupied by these particles volumes and without
overlapping these volumes with ones occupied by another particles in
this space\footnote{Another particles previously located in the being
occupied volume can not stay this volume as they are forced to interact
with the same target the same moment.}. An ordinary week cooling is
observed in the phase space determined in canonical variables. Another
more complicated schemes of external selectivity based on moving
targets must be used for enhanced cooling.

The enhanced longitudinal method of laser cooling based on internal
selectivity in combination with the enhanced transverse method of
damping can be effectively used for three-dimensional cooling of ion
beams (see section IV). An enhanced cooling method based on more
complicated scheme of the external selectivity and moving screens
located on the way of laser targets can be effectively used for
three-dimensional enhanced optical cooling of proton, fully stripped
ion and other particle beams (see section V).

      \section{ENHANCED COOLING OF ION BEAMS
      BEYOND THE ROBINSON'S DAMPING CRITERION.}

Below enhanced laser cooling of ion beams beyond the Robinson's damping
criterion is discussed. Internal ion selectivity in the process of
Rayleigh scattering of laser photons is used. Three examples are
considered.

1) Monochromatic laser beam target with scanning frequency is used when
the RF system of the storage ring is switched off (RF buckets, linear
dependence $\overline {P} ({\Delta \varepsilon })$ and stationary
conditions are absent) \cite {channel}-\cite {hangst3}. The laser beam
overlaps the ion beam. Not fully stripped (electronic transitions) or
naked (nuclear transitions) ions interact with the homogeneous
counter-propagating laser beam at resonance energy, decrease their
energy in the process of the laser frequency scanning until all their
energies reach the minimum energy of ions in the beam. At this
frequency the laser beam is switched off.

The higher the energy of ions, the earlier they start interacting with
the laser beam, the longer the time of interaction. Ions of minimum
energy do not interact with the laser beam at all. At the same time the
amplitude of betatron oscillations is not changed, if the dispersion
function of the storage ring at the IR is zero. As a result, the energy
spread is decreased by nonexponential low to very small value
determined by the width of the ion spectral line, the bandwidth of the
laser beam and the average energy of the emitted photons\footnote
{Exponential damping leads to decrease of the beam dimension $e \simeq
2.7 $ times for one damping time while non-exponential damping leads to
much greater decrease and much faster cooling.}.  The radial amplitudes
and the emittance will not be changed. The damping time is determined
by (\ref {eq7}).

2) Ion and broadband laser beams interact in a straight section of a
storage ring. The laser beam is homogeneous in limits of the ion-laser
beam IR and has sharp frequency edges. The frequency band of the laser
beam is sufficient for all ions to interact with the laser beam.  The
RF system of the storage ring is switched off (violation of conditions
1, 2). The minimum initial energy of ions corresponds to interaction
with laser beam photons of high-frequency edge. The dispersion function
of the storage ring at the IR is zero.

In this case ions decrease their energy until all their energies reach
the minimum energy of ions in the beam, the radial amplitudes and
emittance will not be changed.

3) Ion and broadband laser beams interact in a straight section of a
storage ring. The laser beam is homogeneous in limits of the ion beam
and has sharp frequency edges. The RF system of the storage ring is
switched on. The synchronous energy of ions corresponds to interaction
with photons of high-frequency edge of the laser beam. The spectral
intensity of the laser beam is linearly decreased from a maximum at the
low-frequency edge to zero at the high-frequency edge.

In this case a discontinuity in the rate of energy loss was introduced:
ions with an energy more than the synchronous energy interact with the
laser beam and ions with less energy do not. A synchronous ion does not
lose energy. There is no friction and antidamping of synchrotron
oscillations at the energy $\varepsilon < \varepsilon _s $ and there is
damping at the energy $\varepsilon > \varepsilon _s $ (violation of
conditions 1, 2). The power of the scattered radiation depends on the
ion energy according to the law:  $ \overline {P} = \overline P _{max}
\left[ (\varepsilon - \varepsilon _s ) /\sigma _{\varepsilon,0}
\right]$ at $\varepsilon _s <\varepsilon < \varepsilon _s + \sigma
_{\varepsilon, \,0}$ and $\overline P = 0$ at $\varepsilon <
\varepsilon _s $, $\varepsilon > \varepsilon _s + \sigma
_{\varepsilon,0}$. The minimum damping time will be determined by
(\ref{eq7}) if we accept in (\ref{eq7}) $\overline P _s = \overline P
_{max}$.  The radial amplitudes and emittance will not be changed.

Damping time for enhanced ion cooling (\ref{eq7}) is $\varepsilon / 4
\sigma _{\varepsilon, \,0} > 10 ^2$ times shorter than
damping time for radiative ion cooling (\ref {eq6}) if scattered powers
are the same.

The considered enhanced methods of laser cooling in longitudinal plane
are based on internal resonance selectivity installation-specific for
ions. In combination with the enhanced transverse damping considered in
previous section they can be used effectively for three-dimensional
cooling of ion beams (see Appendix 1) \cite {heacc01}. The emittance
exchange through a synchro-betatron resonance \cite {sessler} or
dispersion coupling by a wedge-shaped laser target can be used for
three-dimensional cooling as well by analogy with the idea of muon
cooling \cite {1983}, \cite {neufer}, \cite {oneil, shoch, sessler1}.

         \section {ENHANCED OPTICAL COOLING OF PARTICLE BEAMS BEYOND
           THE ROBINSON'S DAMPING CRITERION}

Below a method of enhanced optical cooling of particle beams based on
external selectivity is considered. In this method two identical
undulators are installed in different straight sections of a storage ring
with high-dispersion and low-beta functions at a distance determined by a
phase advance $2 p \pi + \pi$ for the lattice segment, where p = 1,2,3...
is a whole number.  Undulator Radiation (UR) emitted by a particle in the
first undulator pass through an optical system with movable screens
located in the image plane of the particle beam. Then this radiation is
amplified and pass through the second undulator together with the
particle. Screens in the optical system open first the way for UR emitted
by particles with higher energies and higher positive deviations from
their closed orbits $x _{\beta _{}} > 0$. The beam of amplified UR in this
case is equivalent to moving prototype of the target $T _{2}$ considered
above if definite phase conditions are fulfilled in the optical system to
inject particles in the second undulator at decelerating phases. In this
case energy losses are accompanied by a decrease both energy spread and
amplitudes of betatron oscillations of particles that is by enhanced
cooling\footnote {If the phase advance for the lattice segment is $2 p \pi
+ \pi$ and the deviation of the particle in the first undulator $x _{\beta
_{}} > 0$, the deviation of the particle in the second undulator $x
_{\beta _{}} < 0$. In this case the amplitude of betatron oscillations of
the particle in the second undulator is decreased.}. After the screen will
open images of all particles of the beam the system must be closed. Then
the cooling process can be repeated. Laser and optical systems in the
considered scheme of cooling are similar to optical systems in the scheme
of the optical stochastic cooling \cite {mikh}.

Another schemes of optical cooling using external selectivity can be
suggested. For example, similar scheme can use first undulator and
even number of undulators installed in straight sections of a storage
ring at distances determined by a phase advance $(2p+1) \pi$ between
neighboring undulators. In this scheme deviations of particles in
undulators "i" and "i+1" are  $x _{\beta _{i}} = - x _{\beta _{i+1}}$ and
that is why the decrease of the energy of particles, according to
(\ref{eq9}), do not lead to change of their betatron amplitudes and
leads to cooling of the particle beam in the longitudinal plane.

The wavelets of UR emitted by a particle in the first undulator and
amplified in the optical amplifier interact efficiently with the
particle in the second undulator and do not disturb trajectories of
another particles if the average distance between particles in
longitudinal direction is less than the length of the UR wavelets $K
\lambda$, where $K$ is the number of the undulator periods; $\lambda$,
the wavelength of the UR\footnote {The amplified UR do not disturb the
next particles (if overlapping occur) in the first approximation and leads 
to a week increasing of their amplitudes of oscillation in the second one 
because of the stochasticity of the phase of the UR wavelet for another 
particles.}.  The efficiency of cooling is higher, if the transverse laser 
beam dimensions of the wavelets in the second undulator are less then the 
transverse dispersion and betatron dimensions of the being cooled particle 
beam.

Enhanced optical cooling in a RF bucket can be produced as well. The
damping time of the order of (\ref{eq7}) can be received.

The considered schemes is of great interest for cooling of proton, muon
and fully stripped ion beams\footnote {Laser cooling based on nuclear
transitions has problems with low-lying levels \cite {ion1}. Optical
cooling of heavy ions, on the level with optical stochastic cooling, is
the most efficient \cite {ion2}. In this case the emitted power $\sim
Z^2$, where $Z$ is the atomic number.}.

              \section*{Appendix 1}

Below we consider the next scheme of the enhanced laser cooling of ion
beams. Moving broadband laser beam target $T_1$ with sharp
high-frequency edge produces damping in the transverse plane.  Then the
target is stopped and produces enhanced cooling in the longitudinal
plane. The RF system of the storage ring is switched off. Non-zero
dispersion function at the IR is used.

Ions having maximal amplitudes of betatron oscillations, according to
(\ref{eq21}), will be deepened in the target after they loose the energy
$\delta \varepsilon /\varepsilon =  \sigma _{x, 0} [\Psi (k _{1,e}) -
\pi \psi (k _{1,e}, \xi _{1,f}) + {1/ k _{1,e}}]/ \eta _x$ necessary
for transverse damping of theirs amplitudes. That is why the frequency
edge of the laser beam must correspond to the minimum initial energy of
ions decreased on the value $\delta \varepsilon$ and the frequency band
of the laser target must be

         \begin{equation}                
             \label{eq28}
         {\Delta \omega _{L_1} \over \omega _{L_1}} \geq {2\sigma
         _{\varepsilon, 0} + \delta \varepsilon  \over \varepsilon}
         = {2\sigma _{\varepsilon, 0} \over \varepsilon}F(R, k _{1,e}),
         \end{equation}
where $F(R, k _{1,e}) = 1 + 0.5 R [\Psi (k _{1,e}) - \pi \psi (k
_{1,e}, \xi _{1,f}) + {1/ k _{1,e}}]$.

In this case the transverse damping time is determined by (\ref{eq24}).
The enhanced cooling in the longitudinal plane is determined by the
damping time

         \begin{equation}                
         \label{eq33}
         \tau _{\varepsilon} = {2\sigma _{\varepsilon, 0} +
         \delta \varepsilon  \over \overline P} = {2\sigma
         _{\varepsilon, 0} \over \overline P}F(R, k _{1,e}).
         \end{equation}

For damping time (\ref{eq33}) ions will be gathered at the minimum
energy corresponding to the high-frequency edge of the laser beam. The
beam will be cooled both in the transverse and longitudinal planes.

              \begin {center} {\bf Example } \end {center}

Below we consider an example for cooling of N-like Xenon ions
($^{129}_{54}Xe ^{47+}$) in the RHIC storage ring when the transition
between the ground state $ (2s^2 2p^3) \,^4S _{3/2}$ and the excited
state $ (2s^2 2p^4)\, ^4P_{3/2}$ is used.  The RF acceleration system
of the RHIC is switched off.  Enhanced ion beam cooling transversely
and fast cooling longitudinally by a homogenous broadband laser with
sharp high-frequency edge is used.

\vbox{            \begin {center}
            \it {Ion properties}
            \end {center}
\vskip -1mm
Degeneracy factors  \hfill  $ g _1 = g _2 = 4$ \\
Transition energy  \hfill  $\hbar \omega _{tr} = 608.44$ eV \\
Wavelength  \hfill  $\lambda_{tr} = 2.04 \cdot 10 ^{-7}$ cm\\
Oscillator strength \hfill   $ f_{12} = 8.9 \times 10^{-2}$ \\
Natural line width  \hfill  $\Delta \omega _{tr} / \omega _{tr} = 1.58
\cdot 10 ^{-6}$ \\
Decay length \hfill $c\tau = 2g _1 f _{1,2}r _e \omega ^2 _{tr}/ g _2
=2$ cm \\

}
\vbox
{            \begin {center}
            \it {Machine parameters}
            \end {center}
\vskip 1mm
Circumference \hfill  $ C = 3.8$ km \\
Relativistic factor  \hfill  $\gamma=97$\\
Energy \hfill   $\varepsilon = 1.18 \cdot 10 ^{13}$ eV \\
Energy spread  \hfill  $2 \sigma _{\varepsilon, \,0} = 4.72\times
10^{9}$ eV\\
Relative energy spread  \hfill  $2\sigma _{\varepsilon, \,0}/{
\varepsilon} = 4 \times 10^{-4}$\\
Emittance  \hfill  $\epsilon_x = 4 \times10 ^{-7}$ m-rad \\
$\beta-$function \hfill   $\beta _x = 1$ m \\
Dispersion function at the IR \hfill $\eta _x = 5$ m \\
RMS dispersion beam size \hfill $2\sigma _{\eta, 0} = 2$ mm \\
RMS betatron beam size \hfill $\sigma _{x,0} = 0.632$ mm
}
\vbox{
            \begin {center}
            \it {Parameters of the laser beam $T_1$}
            \end {center}
\vskip 1mm
Wavelength  \hfill  $\lambda_{L_1} = 3954$ \AA\\
Intensity \hfill $I _{L_1} = 1.16 \times 10 ^6$ W/cm$^2$ \\
Saturation parameter \hfill          $D =10 ^{-2}$ \\
Bandwidth  \hfill  $\Delta \omega _{L_1} / \omega _{L_1} = 9.7 \cdot 10
^{-4}$\\
Intraresonator power \hfill  $P _{L_1} = 2 \pi \sigma _{L_1} ^2 I _{L_1}
\simeq 0.45 $ MW \\
RMS radial beam size \hfill   $\sigma _{L _1} = 2.5$ mm    \\
Rayleigh length \hfill   $z _R = 4 \pi \sigma _{L_1} ^2 / \lambda
_{L_1} \simeq 20.8$ m \\
Interaction length \hfill $l _{int} = 10$ m }

\vskip 4mm
The required power of the laser beam should be feasible to obtain with
a free electron laser in an intracavity configuration \cite
{prl}\footnote{ At present such lasers are in operation \cite {benson},
\cite {minehara}.} or with a conventional laser using a high-finesse
($\sim 10 ^4$) optical resonator \cite {zh}\footnote {Finesse of $1.9
\cdot 10 ^6$ was reported near $\lambda = 850$ nm in \cite {rempe}.}.

In this case the saturation intensity $I_{sat} = [g _1/(g _1 + g _2)](2
\pi c \hbar \omega _{tr}/\gamma ^2 \lambda _{tr} ^3)$ $(\Delta \omega
_L/\omega _L) \simeq 2.87 \times 10 ^8$ W/cm$^2$; the power of the
radiation scattered by ion $\overline P = 2\gamma ^2 l _{int} P _L
\overline \sigma _R/ (1 + D) C S _{eff}$ $= 8.79 \cdot 10 ^{-10}$ W $=
5.49 \cdot 10 ^{9}$ eV/sec $\simeq 6.95 \cdot 10 ^4$ eV per turn; the
effective area of the interaction region of laser and ion beams $S
_{eff} = 2 \pi (\sigma _L ^2 + \sigma _{x,0} ^2) = 0.42$ cm$^2$; the
Rayleigh scattering cross-section $\overline \sigma _R = \pi f _{12} r
_e \lambda _{tr} \omega _L / \Delta \omega _L = 6.434 \cdot 10 ^{18}$
cm$^2$; $r _e = e ^2 / mc ^2$ \cite {prl}, $|k_1| = 0.1565$, $R =
0.632$, $F(R, k _{1,e}) = 2.36$. The bandwidth of the laser beam is in
accordance with (\ref{eq28}).

For these parameters of ion and laser beams, the velocity of the closed
orbit of an ion in the laser beam $\dot x _{\eta \,in} = \eta _x
\overline P/ \varepsilon = 0.233$ cm/sec, the average energy of the
scattered photons is $<\hbar \omega^s> = \hbar \omega _{tr} \gamma =
59.0$ keV, the velocity of the laser beam $v _{T_1} = 3.64 \cdot 10
^{-2}$ cm/sec, $\sigma _{x,f} = 0.232$ mm, $2\sigma _{\varepsilon,f} =
\varepsilon (\Delta \omega _L/\omega _L) = 1.14 \cdot 10 ^{10}$ eV,
$2\sigma _{\eta, f} =4.85$ mm, $2\sigma _{f} < a = 2\sigma _{L _1}$.

In this case, according to (\ref{eq24}), the transverse damping time
$\tau _x = 5.5$ sec. and, according to (\ref{eq33}) the longitudinal
damping time $\tau _{\varepsilon} \simeq 2.03$ sec. The damping time of
the same beam in the method of radiative ion cooling, according to
(\ref{eq6}), is $\tau _x = \tau _y \simeq \tau _{\varepsilon} \simeq
4.3 \cdot 10 ^3$ sec, i.e., $\sim 10 ^{3}$ times higher\footnote{In
this scheme a broadband laser beam overlaps an ion beam, all ions
interact with the laser beam independently of their energy and
amplitude of betatron oscillations \cite {prl}, \cite {idea}-\cite
{pac}. The physics of radiative ion cooling is similar to SR damping.}.

The degree of the reduction in the dispersion beam radius in the
non-exponentional longitudinal cooling of the ion beam can be very
small ($\sigma _{\eta,\,f}  /\sigma _{\eta,\,0}  \ll 1$). It depends on
the sharpness of the high frequency edge of the laser beam and quantum
processes\footnote {The energy spread of the being cooled ion beam is
limited by quantum processes of Rayleigh scattering to the value
$\sigma _{\varepsilon, q} \simeq$ $<\hbar \omega ^s> = 59$ keV.}.

Cooling of the beam in the longitudinal plane can be performed by a
monochromatic laser beam with a scanning frequency for the time
determined by (\ref{eq33}), which is $\tau _{\varepsilon} \ll \tau _x$.
It can be located in another straight section of the storage ring.
Unfortunately, using of the optical resonator in this case (laser beam
with a scanning frequency) has problems.

To increase the degree of damping of betatron oscillations in the
transverse plane we can use a target moving back and forth between
outside orbit of the beam to inside one. A phase displacement mechanism
or eddy electric fields can be used for the recovery of the ion beam
energy and production of the next cooling cycle.

Schemes of six-dimensional cooling in the RF bucket can be suggested as
well.

In the considered method of ion cooling, the laser beam does not have a
sharp edge. Since the intensity of the beam is changed with the
displacement of the ion orbits to the laser beam center, the laser beam
will be similar to the wedge-shaped target. Such targets at rest
decrease the amplitudes of ion betatron oscillations in case of the
target $T _1$ and increase them in case of the target $T _2$ in lesser
degree than the targets with sharp edges. This means that the
requirements for sharpness of the laser target $T _1$ are not strong.
A system of lasers can be used to increase the sharpness of the
effective composite laser beam. For example, a third laser beam of the
same maximum intensity as the first one with the dimension $\sigma _{L
_3} \simeq \sigma _{L _{1}}/4$ intersecting the ion beam at a third
straight section of the storage ring with the displacement $2 \sigma
_{L _1}$ relative to the radial position of the first laser can be used
in the example.

Cooled not fully stripped ion beams can be used in New Generation Light
Sources \cite {esrf96}.

                    \section {Conclusion}

We investigated the limits of applicability of the Robinson's damping
criterion to the problem of cooling of particle beams in storage rings.
Theory of the emittance exchange based on external selectivity caused by
moving material targets is developed. New schemes of six-dimensional
enhanced cooling of ion beams based on internal selectivity of ions beyond
the criterion are considered. A scheme of enhanced optical cooling of
particle beams (proton, ion, muon) based on external selectivity is
suggested and developed.

Author thanks A.M.Sessler and Robert Palmer for useful discussion and
acknowledges the support of this work by the RFBR under Grant No
02-02-16209.

\newpage
\end{document}